\documentclass[conference]{IEEEtran}
\ifCLASSINFOpdf
  \usepackage[pdftex]{graphicx}
\else
\fi

\usepackage{hyperref}
\usepackage{subfig}

\usepackage[normalem]{ulem} 
\usepackage{xcolor}

\usepackage{ifthen}
\usepackage{amssymb}
\newboolean{showcomments}
\setboolean{showcomments}{true} 
\ifthenelse{\boolean{showcomments}}
 {\newcommand{\Note}[3]{\fcolorbox{gray}{#3}{\bfseries\sffamily\scriptsize#1}{\sf\small$\blacktriangleright$\textit{#2}$\blacktriangleleft$}}}
  {\newcommand{\Note}[3]{}}

\usepackage{alltt}
\begin{document}
%
\title{On Extracting Unit Tests from\\Interactive Programming Sessions}

\author{\IEEEauthorblockN{Adrian Kuhn}
\IEEEauthorblockA{Department of Computer Science\\
University of British Columbia\\
Twitter: \href{http://twitter.com/akuhn}{@akuhn}}}


%


\maketitle

\begin{abstract}
Software engineering methodologies propose that developers should capture their efforts in ensuring that programs run correctly in repeatable and automated artifacts, such as unit tests. However, when looking at developer activities on a spectrum from exploratory testing to scripted testing we find that many engineering activities include bursts of exploratory testing. In this paper we propose to leverage these exploratory testing bursts by automatically extracting scripted tests from a recording of these sessions. In order to do so, we wiretap the development environment so we can record all program input, all user-issued functions calls, and all program output of an exploratory testing session. We propose to then use machine learning (i.e. clustering) to extract scripted test cases from these recordings in real-time. We outline two early-stage prototypes, one for a static and one for a dynamic language. And we outline how this idea fits into the bigger research direction of programming by example.

\end{abstract}


%
\IEEEpeerreviewmaketitle

\section{Introduction}

Software engineering methodologies propose that developers should capture their efforts in ensuring that programs run correctly in repeatable and automated artifacts, such as unit tests \cite{Beck2002a}. However, when looking at developer activities on a spectrum from exploratory to scripted testing we find that many engineering activities include bursts of exploratory testing \cite{Whit2009a}, for example when developers execute selected features of a program on the REPL\footnote{REPL, \emph{abbr.\ for Read Evaluate Print Loop}: interactive live programming interface that is commonly found in the toolbox of dynamic languages. Similar tools exist for static languages in development environments like Eclipse (Java) and VisualStudio (C\#), but other than REPLs they tend to be rarely used by practitioners from the respective language communities.} to ensure some sample input yields correct results, but also when engineers instrument programs with print statements during debugging sessions. 

In this paper we propose to leverage these exploratory testing efforts by automatically extracting scripted tests from a recording of these sessions. We call our approach REX, which is short for REPL Extraction. In order to do so, we plan to wiretap the development environment so we can record all program input, all user-issued functions calls, and all program output of an exploratory testing session. We propose to then use machine learning (i.e. clustering) to extract scripted test cases from these recordings in real-time.

The remainder of the paper is structured as follows. 
In \autoref{sec_example} we provide a motivating example. 
In \autoref{sec_related} we discuss related work. 
In \autoref{sec_approach} we outline the proposed approach and two proof-of-concept prototypes. 
In \autoref{sec_recap} we close with concluding remarks.

\section{REPL Extraction in a Nutshell}
\label{sec_example}

In this section we provide a motivating example of test case extraction. Below we show the log file of an interactive programming session that explores Ruby's rational number API. The sessions explores whether adding two numbers returns a reduced representation of the result (user input is set in bold for better readability):

\begin{alltt}
\small
Macchiato:nier2013 akuhn\$ \textbf{irb}
>> \textbf{require 'rational'}
=> true
>> \textbf{Rational.new(1,3)}
NoMethodError: private method `new' called for 
Rational:Class from (irb):5
>> \textbf{Rational(1,3)}
=> Rational(1, 3)
>> \textbf{_ + Rational(1,6)}
=> Rational(1, 2)
>> \textbf{quit}
\end{alltt}

From which the REX algorithm would extract and generate the following RSpec\footnote{http://rspec.org} test case (keywords of the RSpec framework are set in bold font for better readability):

\begin{alltt}
\small
require 'rational'
\textbf{describe} Rational \textbf{do}
  \textbf{it} "should +" \textbf{do}
    x = Rational(1,3)
    y = x + Rational(1,6)
    y.inspect.\textbf{should match} "Rational(1, 2)"
  \textbf{end}
\textbf{end}
\end{alltt}

To generate the test case, the recorded session is partitioned into three phases: setup (the require statement), interaction (creating the two rational numbers) and assertion (checking that the final print statement is as recorded). Also some filtering happens: Illegal operations are common in interactive sessions and we thus filter them out. In the same line, intermediate output has been discard and only the final output used in the \texttt{\textbf{should match}} assertion. For a discussion of how to accommodate for the possibility that illegal operations or intermediate output are actually the developer's intended assertion under test, please refer to \autoref{sec_approach}.

\section{Related Work}
\label{sec_related}

There is a rich body of research on test case and test data generation. However most of this work deals with generating test through program analysis, or by using record-and-replay of interactions at the level of the program's user interface. For a good coverage of test generation the reader is referred to the proceedings of the international conference on Automated Software Engineering (ASE). However, to our best knowledge there is little to no work on automatically extracting unit tests from interactive REPL sessions. 

More closely related to the ideas of this paper is work on \emph{programming by example}, as for example Edward's work on example-centric programming \cite{Edwa2004a}. He presents a development environment with a two editor view: one view shows the stack trace of the current program and the other view shows the source code. Both views can be edited and update one another. And also related is the work by Brad Myers et.al. where they built systems in which the user defines behaviors by demonstration \cite{Myer2000a}; in some of them, the system uses AI algorithms, such as inferring complex behavior from a few examples; in others, the user has to provide the full specification, and the examples are used primarily to help the user understand the programming situation. 

Smalltalk and LISP, as well as their modern incarnations Pharo\footnote{http://pharoproject.org} and Clojure\footnote{http://clojure} are also related work, at least in the spirit of standing on the shoulder of giants, as they pioneered the idea of live programming on which this paper relies.
%

\section{Vision and Approach}
\label{sec_approach}

With this paper we not only outline a novel tooling approach but also a novel development methodology, inspired by the idea of ``tests as examples'' \cite{Gael06b}. Example-driven development regards unit tests not just as artifacts of scripted testing, but also as documentation and as resumable program state. Documentation in the sense that the source code of a test captures idiomatic use of the unit under test, and resumable program state in the sense that a test's return value can serve as the initial state for either a live programming session or as the setup of another test. 

\begin{figure}[t!]
\centering
\includegraphics[width=\columnwidth]{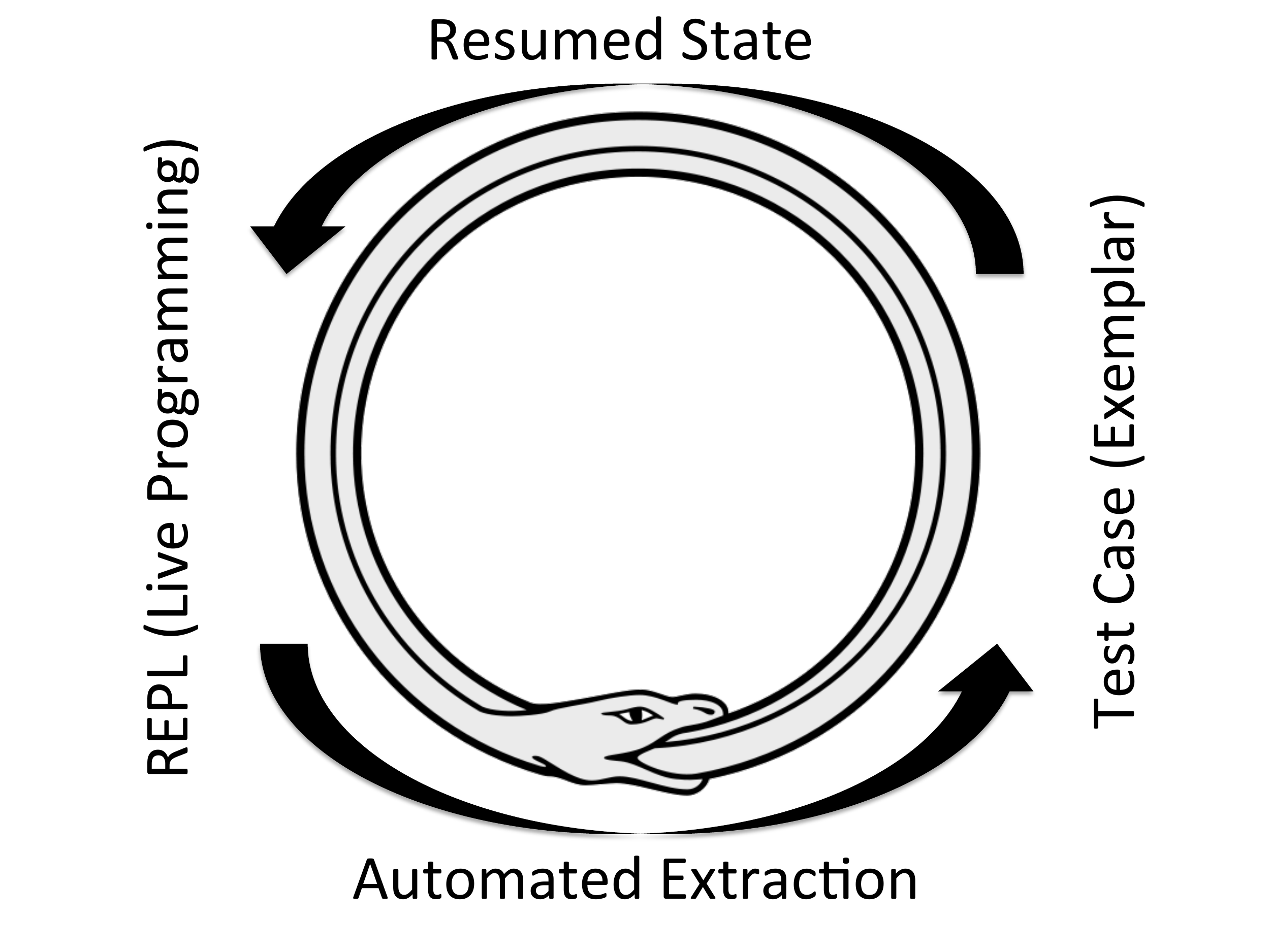}
\caption{Illustration of example-driven development in a REX-enable environment: On the left are REPL sessions on the right test cases, here referred to as examples. REPL session can be started using a test's return value as initial state (top arc of the circle) and the present REX approach extract test cases from  REPL session (bottom arc of the circle).}
\label{fig_orborous}
\end{figure}

\autoref{fig_orborous} illustrates example-driven development: On the left are REPL sessions on the right test cases, here referred to as examples. REPL sessions can be started using a test's return value as initial state (top of the circle) and the proposed REX approach extracts test cases from  REPL sessions (bottom of the circle). Thus the circle closes.

We conjecture that in a REX-enabled develop environment most of the programming happens in interactive sessions that are started by selecting from existing examples, similar to development practices in Smalltalk where most programming is known to happen in the debugger rather then the editor. In particular once we extend the REX approach to extract not just test cases but also functions. Such a system would eventually enable users to program the computer by providing example data and doing example manipulations on this data, from which a computer would infer the program logic. 

This is a far reaching goal, hence the focus of this paper is on the design of a system that extracts test cases form example interactions during a live programming session. We hope that a system like that will support and encourage practitioners in industry in curating better and more complete test suites of their systems. In particular in industries with a ``big-P'' process where the role of writing software and the role of testing software are possibly assigned to separate people. The capability to capture the exploratory activity of engineerings roles and passing it on as a starting point to testing roles could bring considerable benefits to such settings.

The design space of test case extraction from programming sessions is large. For a first exploration of the design space we decided to go for languages with interactive REPL environment (think Ruby, Javascript tools in modern web browsers, but also Eclipse's Scrapbook and Display views) as it is easy to acquire a transcript of these sessions. We imagine that future designs could also, for example, listen to print statements that developers add during debugging sessions or even record the developer's interaction with a GUI-based debugger and then extract test cases from these recordings. In the following, we are going to describe two proof-of-concept prototypes that aim for the most simple thing that could possibly work, one for the Ruby language and one for the Java language.

\subsection{REX Prototype for \texttt{irb}, the Interactive Ruby Console}

The interactive Ruby console, \texttt{irb}, as shown in \autoref{fig_irb}, is a classic read-evaluate-print loop that prompts for user input, evaluates the input and prints the result. Our current prototype works in three steps 1) capturing of the REPL's transcript, 2) clustering of the transcript into a tree of hierarchical test cases, 3) omitting the source code of an RSpec\footnote{http://rspec.org} test suite.

At the moment we capture a transcript of the interactive sessions by manually saving a dump of the terminal session. We customized the prompt of \texttt{irb} to include the current system time. These timestamps are used to partition the transcript into bursts of exploratory testing. We use hierarchical clustering and cut it off at a threshold of 90 seconds. This is based on the assumption that exploratory testing happens in bursts where for during exploration session no two instruction are separated by more than 90 seconds, while the sessions themselves are separated by more then 90 seconds. We further group the session by dependencies to global variables (when using a REPL global variables are used to store state). 

The previous step results in a hierarchical tree of partitioned REPL instructions. We then turn this tree into an RSpec test suite of the following format:

\begin{figure}[t!]
\centering
\includegraphics[width=\columnwidth]{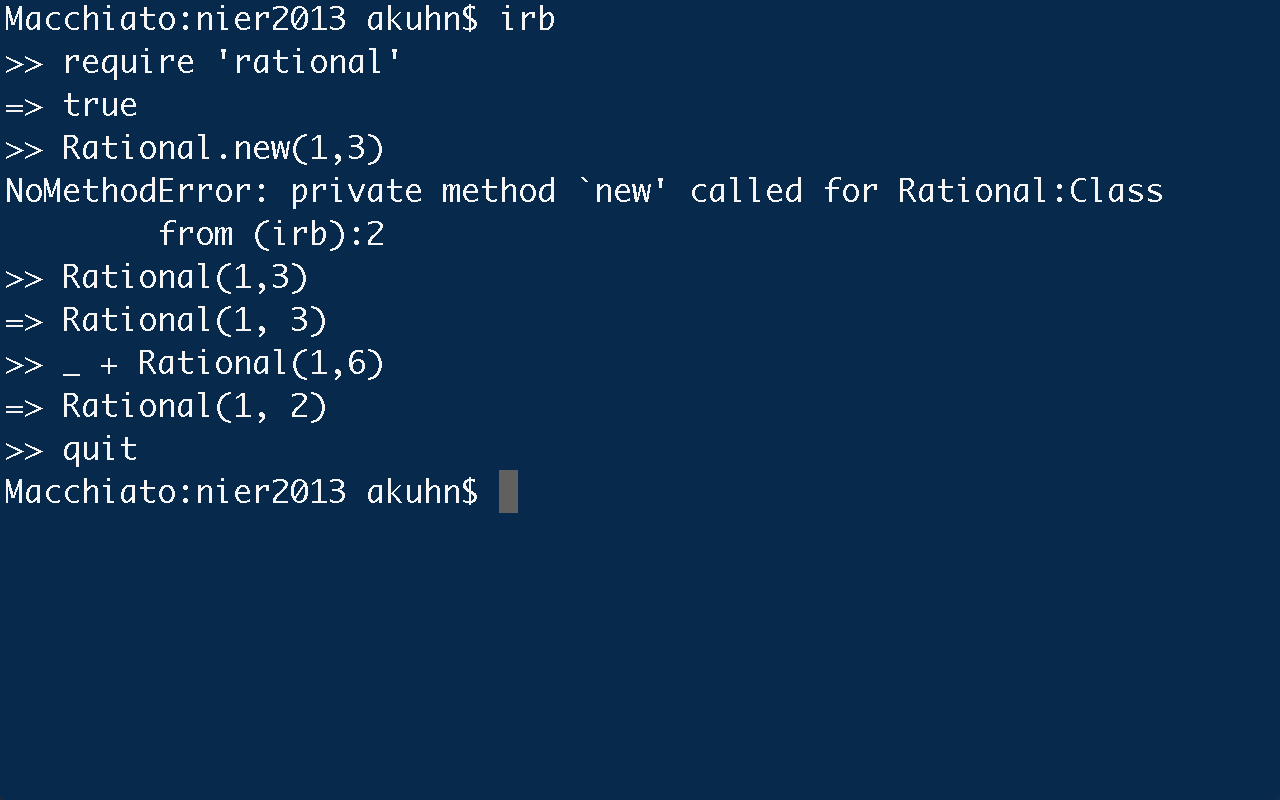}
\caption{Screenshot of \texttt{irb}, Ruby's interactive console}
\label{fig_irb}
\end{figure}

\begin{alltt}
\small
\textbf{[extracted imports]}
describe \textbf{[class under test]} do
  before :each do
    \textbf{[extracted setup]}
  end
  it \textbf{[summary of methods under test]} do
    \textbf{[extracted interaction]}
    x.to_s.should match \textbf{[extracted output]}
  end
end  
\end{alltt}  

RSpec allows test cases and their setup to be grouped in nested contexts, we can thus transform the tree of extracted test cases without having to flatten it down. So for example, assumed all test cases depend on variable \texttt{a} and some depend on either variable \texttt{b} or variable \texttt{c}, then all test cases would be grouped in one \texttt{describe} context that sets up the initialization of variable \texttt{a}, and nested within that context would be a context each for variable \texttt{b} and variable \texttt{c}. Individual test cases are put in \texttt{it} blocks. Using nested contexts is encouraged best practice when using RSpec \cite{Chel2001a}.

When emitting source code for the test suite, instructions that have lead into an error are excluded. We do not exclude them from the previous clustering step as they might still provide hints that are useful to group the test cases by variable dependency. For each test case we emit a \texttt{should match} assertion for the textual output of the final command. This is based on the assumption that engineers finish an exploratory testing session after having seen a correct output. 

\begin{figure}[t!]
\centering
\includegraphics[width=\columnwidth]{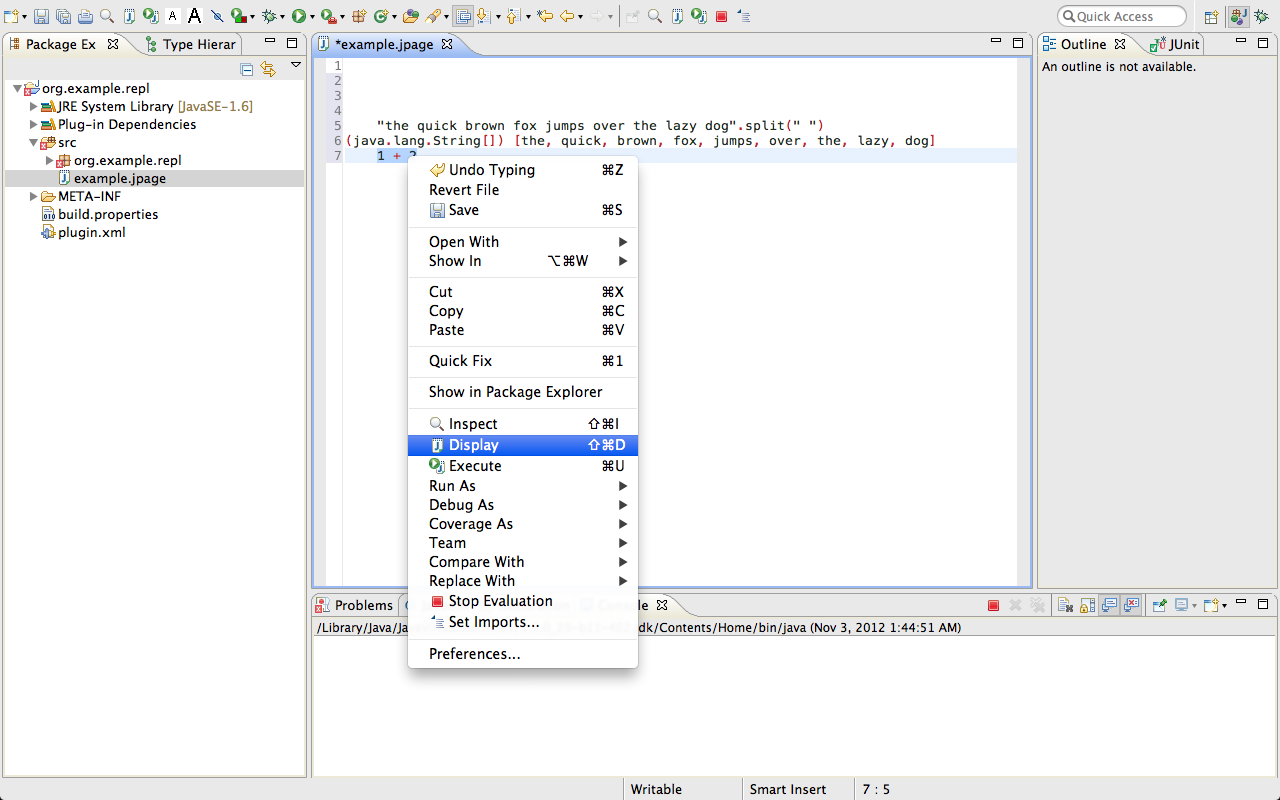}
\caption{Screenshot of Eclipse's interactive \emph{Scrapbook} view.}
\label{fig_Scrapbook}
\end{figure}

If the last instruction resulted in an error we initially used to add an assertion that checks for the presence of this error, but we found that these tests often tended to fail when running the test suite so we currently do not emit any assertion for these test cases. The value of test cases without assertion is debatable, but they might still serve as smoke tests.

Obviously, asserting against the printed representation of an object is fragile with regard to changes to its print method (for Ruby \texttt{to\_s} and \texttt{inspect}, for Java \texttt{toString}), however we conjecture that print method change less frequently than other methods of an object. This conjecture is based on a recent study on the use of print methods in object-oriented systems \cite{Schw11b}. The idea of striving for simplicity and testing a system by its output rather than deep inspection is also applied in the Sikuli tool, which matches selected screenshots against the current display on the screen to test user interfaces \cite{Yeh09a}.

While dogfooding\footnote{Dogfooding refers to using your own tools to develop them.} our prototype we found several gotchas to be covered when matching against textual object representations. For example, some representations (e.g. Ruby's default \texttt{Object\#inspect} method) include hash values that differ between program runs. For the same reason, ordering of sets/hash or any data that has been passed through a set or hash might be different. Time might be different. If live data is fetched from the web, that might be different. One possible strategy to alleviate this is to rerun the recorded transcript, and then use regular expressions that exclude those letters that have differed between the original and the repeated REPL sessions. So for example:

\begin{alltt}
\small
Macchiato:nier2013 akuhn\$ irb
>> Object.new
=> \#<Object:0x106a2a628>
>> quit 
Macchiato:nier2013 akuhn\$ irb
>> Object.new
=> \#<Object:0x2119c85e0>
>> quit 
\end{alltt}

would result in:

\begin{alltt}
\small
describe Object do
  it "should new" do
    x = Object.new
    x.inspect.should match /\#<Object:0x(.*)>/ 
  end
end   
\end{alltt}

To summarize test cases and contexts we use a very simple approach. For contexts we name them after all class names (i.e. names starting with an uppercase letter) that appear in their body. And for test cases we use all methods names that appear in their body and prepend them with the verb ``should'' as it is convention for RSpec tests. There is work on more elaborate test case summarization \cite{Kami2012x} that we could leverage here. 

\subsection{REX Prototype for Eclipse's Scrapbook view}

The Eclipse debugger features an interactive programming tool as well, the \emph{Scrapbook} view, shown in \autoref{fig_Scrapbook}. This view is akin to Smalltalk's workspaces, that is a text editor where programmers can evaluate selected text and the results of the evaluation is inserted into the editor after the end of the selection. Again as with the Ruby prototype, our prototype works in three steps 1) capturing of the Scrapbook's content, 2) clustering of the content into a tree of hierarchical test cases (not yet implemented due to limitations of the Scrapbook interface, see discussion below), 3) emitting the source code of a JExample\footnote{http://scg.unibe.ch/jexample} test suite.

To accommodate for the lack of a linear reading sequence in Eclipse's Scrapbook view, our current prototype is limited to coding sessions where all instruction are written one below the other, and all evaluation output is on separate lines. We found that these constraints do not limit the usefulness of the Scrapbook for interactive programming. The content of the Scrapbook is then automatically captured by our REX plug-in. We do not currently cluster the content automatically due to limitation of the  Scrapbook view, which neither offers a customizable prompt nor global variables. For future work we intend to write our own interactive Java REPL based on the scrapbook's internal interpreter that would be free of such limitations; and even allow us to extract tests in real-time as we are getting a real-time interaction signal.

Eventually we emit source code for a JExample test suite. At the moment these test suites are minimal due to the limitation of our current Java tooling, however in the future REX will allow us to emit source code to hierarchical test suites. JExample allows test cases to be composed into a dependency tree, and offer the option to reuse a test's return value as fixture for dependent tests \cite{Kuhn08a}.

While the current Java prototype is not impressive, it proves that interactive coding sessions are possible in static languages and that there is the potential to write an interactive REPL for Java using the internal interpreter of Eclipse\footnote{For the interested reader and prospective Eclipse plug-in developer, the interpreter is found in \texttt{org.eclipse.jdt.internal.debug.ui.eval} and works without requiring an active debugging session to another JVM.}.

\section{Concluding Remark}
\label{sec_recap}


In this paper we presented an approach to extract unit tests from interactive programming sessions. We outlined our ideas for this line of research, which we hope will eventually lead to a programming-by-example system where not only test cases but also functions are extracted from live programming sessions. We presented two prototypes, one in Ruby and one in Java, as a proof of concept of our idea. We would like to extend these prototypes in future work to improve their stability, applicability and usefulness. In the near future we hope to provide benefits to developers in both engineering and testing roles, by providing them with means to automatically capturing their exploratory testing efforts as scripted tests.

\bibliographystyle{IEEEtran}
\bibliography{replmining-icse2013.bib}

\end{document}